\documentclass[aps,nofootinbib]{revtex4}

\usepackage{amsmath}
\usepackage{amssymb}
\usepackage{bbm}
\usepackage{mathtools}
\usepackage[normalem]{ulem}
\usepackage{dsfont}
\usepackage{mathrsfs}
\usepackage[makeroom]{cancel}
\usepackage{graphicx}
\usepackage{bm}%ważne: aby tilde nad bold symbolami bylo OK

\usepackage[ddmmyyyy]{datetime}
\makeatletter
\def\Dated@name{}
\makeatother

\def\rhat{{\hat {\boldsymbol r}}}

\def\0{\boldsymbol{0}}
\def\RR{\mathbbm{R}}

\def\BD#1{{\boldsymbol{#1}}}

\def\V{\boldsymbol{V}}

\def\Lag{{\cal L}}
\def\ii{\mathrm{i}}

\def\bnabla{{\boldsymbol\nabla}}

\def\VAREPS#1{\varepsilon_{#1}}
\def\OM#1{\omega_{#1}}

\def\B#1{\left(#1\right)}
\def\BB#1{\left[#1\right]}
\def\BBB#1{\left|#1\right|}

\def\be{\begin{equation}}
\def\ee{\end{equation}}

\def\for{\ \text{for} \ }

\def\pref{0.6}

\begin{document}
\title{Reference frame dependence of the periodically oscillating 
Coulomb field in the Proca theory}
\author{Bogdan Damski}
\affiliation{Jagiellonian University, 
Faculty of Physics, Astronomy and Applied Computer Science,
{\L}ojasiewicza 11, 30-348 Krak\'ow, Poland}
\begin{abstract}
The Proca theory of the real massive vector field 
admits non-equilibrium solutions, where the 
 asymptotic dynamics of its electric field 
is dominated by the periodically oscillating 
Coulomb component. We discuss  how such 
field configurations are seen in different reference frames,
where we find an intriguing spatial pattern of the vector field
of the Proca theory 
and the  electromagnetic field associated with it.
%\\ \\
%\version{frame31} 
\end{abstract}
%\date{\today}
\maketitle

\section{Introduction}
\label{Introduction_sec}
 
We consider the simplest version of the 
Proca theory, which is defined by the following
Lagrangian density 
\be
\Lag=-\frac{1}{4} 
(\partial_\mu V_\nu-\partial_\nu V_\mu)^2
+\frac{m^2}{2}(V_\mu)^2,
\label{SPr}
\ee
where 
 $V$ is the vector field  
 and $m>0$ is the mass of the vector 
boson 
(see Appendix \ref{Conv_app}  for our conventions).
This  theory is often used  to illustrate 
various  field theoretic issues
and it describes some properties of massive 
vector bosons such as $\rho$ and  $\omega$ mesons \cite{Greiner,ColemanBook,Weinberg}.
Moreover, the Proca theory is considered as
an  
extension of Maxwell electrodynamics \cite{Nieto_RMP1971,Gillies2005,Nieto_RMP2010}, 
the one in which  a photon is   a massive particle.
Various extensions  of the Proca theory 
are 
discussed in a cosmological context in  \cite{HeisenbergPhysRep2019}.

We recently studied the quantum version of the Proca theory
in \cite{BDPeriodic1}, where the periodically oscillating 
Coulomb field was introduced  for the first time.
In this work,
we focus entirely on the classical version of the Proca
theory,
where the reference frame dependence of such an
unusual   field can be most transparently discussed.

To proceed, we introduce the primed reference frame, 
where
the vector field is denoted as $V'$ and the spacetime
coordinates are $(t',\BD{r}')$, where $\BD{r}'=(x',y',z')$.
Then, we consider the following 
field configuration
\begin{subequations}
\begin{align}
\label{Vprim0}
&V'^0(t',\BD{r}')=-\frac{q}{m^2}\int
\frac{d^3k}{(2\pi)^3}f(\OM{k})\exp(\ii\BD{k}\cdot\BD{r}')\cos(\VAREPS{k}t'),\\
&\V'(t',\BD{r}')=\frac{\ii q}{m^2}\int
\frac{d^3k}{(2\pi)^3}f(\OM{k})\frac{\VAREPS{k}}{\OM{k}^2}
\BD{k}\exp(\ii\BD{k}\cdot\BD{r}')\sin(\VAREPS{k}t'),
\label{Vprimvec}
\end{align}
\label{Vprim}%
\end{subequations}
where $q$ is a constant non-zero parameter,
$\OM{k}=|\BD{k}|$, $\VAREPS{k}=\sqrt{m^2+\OM{k}^2}$,
and real-valued  $f(\OM{k})$  weights 
the Fourier modes in the studied field configuration
($f(0)\neq0$ is assumed  because low frequency modes are of  interest in 
this work).
We assume that $V'$ represents the finite-energy 
field configuration and that its first and second order 
derivatives over space-time coordinates can be obtained
by taking such derivatives under the integral symbol,
which facilitates the  computation of the electromagnetic field
of the Proca theory
and the verification of Proca field equations.
These properties can be  ensured by various functions  
$f(\OM{k})$, e.g. such as those given by  (\ref{fbeta}).

Taking into account the above remarks, 
one may easily verify that 
  (\ref{Vprim}) satisfies 
the Proca field equations in the 
primed reference frame
\be
\Box'V'=-m^2V', \ \partial' \cdot V'=0,
\label{ProcaF}
\ee
where $\Box'=\partial'\cdot\partial'=\partial_{t'}^2-
\partial_{x'}^2-\partial_{y'}^2-\partial_{z'}^2$.
Moreover, via 
$\BD{E}'=-\partial_{t'}\V'-\bnabla' V'^0$ and $\BD{B}'=\bnabla' \times \V'$,
where $\bnabla'=(\partial_{x'},\partial_{y'},\partial_{z'})$,
one may also verify that 
\begin{align}
\label{Eprim}
&\BD{E}'(t',\BD{r}')=-\ii q\int
\frac{d^3k}{(2\pi)^3}f(\OM{k})\frac{\BD{k}}{\OM{k}^2}
\exp(\ii\BD{k}\cdot\BD{r}')\cos(\VAREPS{k}t') ,\\
&\BD{B}'(t',\BD{r}')= \0.
\label{Bprim}
\end{align}
 $\BD{E}$ and $\BD{B}$ are defined (in terms of the 
vector field) in the same way as the electric 
and magnetic fields
in the Maxwell theory. Thereby, for the sake of brevity, 
we will refer to them as the electric and magnetic fields 
in our  Proca theory studies 
despite the fact that the vector field involved in their 
definition describes a massive spin-1 boson  (not a massless 
one as is the case in the Maxwell theory).
This important fact should be kept in mind while reading this work.

The discussed field configuration
is described by 
spherically symmetric 
$V'^0(t',\BD{r}')$ as well as
$\BD{V}'(t',\BD{r}')\propto\BD{r}'$
and 
$\BD{E}'(t',\BD{r}')\propto\BD{r}'$,
where the proportionality factors 
are  a function of $t'$ and 
 $r'=|\BD{r}'|$.\footnote{These 
      statements 
      can be  justified  by doing the
angular integration in 
(\ref{Vprim}) and (\ref{Eprim}). 
Writing 
$d^3k$ as 
$\mathit{d\OM{k}}\OM{k}^2\mathit{d\Omega(\hat{\BD{k}})}$,
this  amounts to 
the evaluation  of 
$\int \mathit{d\Omega(\hat{\BD{k}})}$
in the above-mentioned expressions.
}
Thereby, our field configuration
is centered around $\BD{r}'=\0$ for 
all times 
and as such it  is at rest in
the primed reference frame. 
As we  deal here with the non-equilibrium 
solution of (\ref{ProcaF}),
the term ``at rest'' refers to 
the fact that the 
analyzed field configuration 
does not move as  a whole  in any particular direction.
In addition to that, we note that 
the energy of the studied field configuration
is given by 
\be
\frac{q^2}{4\pi^2 m^2} \int_0^\infty
\mathit{d\OM{k}} f^2(\OM{k})\VAREPS{k}^2.
\ee
This result has 
been obtained from the standard expression
\cite{Greiner}
\be
\frac{1}{2}\int d^3r' \left(
|\BD{E}'(t',\BD{r}')|^2 + |\BD{B}'(t',\BD{r}')|^2   +m^2|\V'(t',\BD{r}')|^2 +  m^2[V'^0(t',\BD{r}')]^2\right),
\label{Eproc}
\ee
 where the presence of the vector field  
does not cause any problems because
such a quantity is observable  in the Proca theory 
(e.g.  \cite{LakesPRL1998} 
reports an attempt of the  experimental
measurement of the vector field  of the Proca theory;
see the discussion by the end of this section).

Our interest in  (\ref{Vprim})
comes from the fact that the following quantity 
exhibits  intriguing  dynamics
\be
Q'(t',\RR^3)=\int  d^3r' \bnabla'\cdot\BD{E}'(t',\BD{r}')=q\cos(mt'),
\label{Q1}
\ee
which has been obtained by employing  (\ref{Eprim})
and the fact that without loss of generality we
assume    $f(0)=1$  (the latter
can be seen as a
normalization condition imposed on $f(\OM{k})$
considered in this work).
Importantly, apart from reliance on  $f(0)=1$, 
(\ref{Q1})
does not explicitly depend on the details 
of $f(\OM{k})$.
Adopting the nomenclature introduced in
\cite{Nieto_RMP1971,Nieto_RMP2010},
one  may call (\ref{Q1})  a
spurious charge or a pseudocharge.
We will call it simply the charge following 
the nomenclature employed in \cite{BDPeriodic1}.
One should, however, keep in mind that we 
talk here  about the charge associated with the
 internal Proca $4$-current 
$-m^2 V'=(\bnabla'\cdot\BD{E}',-m^2\BD{V}')$
 (not the charge associated with  
some external $4$-current  describing electrically charged 
particles). We mention in passing  that 
(\ref{Q1}) illustrates the phenomenon of
periodic charge oscillations in the classical 
Proca theory. Such a phenomenon 
was  recently  discussed
in the quantum Proca theory  \cite{BDPeriodic1},
where it follows from the harmonic 
oscillator equation satisfied by the 
charge operator  \cite{GHK1964,Guralnik1968}.
A similar phenomenon was mentioned  in a different 
physical context in \cite{Hertzberg2020}.

The importance of 
(\ref{Q1}) lies in the fact that such a result, 
via the Gauss' law argument, 
 shows  that
vector field 
(\ref{Vprim}) describes some sort 
of the Coulomb field in the limit of 
$r'\to\infty$:
the one exhibiting both the inverse-square law
 decay and  oscillations in the 
time domain  with the period $\tau'=2\pi/m$.
The reason  why it is so 
is the following: the Fourier
transform
$\int d^3r' \BD{E}'(t',\BD{r}')
\exp(-\ii\BD{k}\cdot\BD{r}')$ 
for   $\OM{k}\to0$
is the same as the Fourier 
transform of the 
Coulomb field multiplied by 
 $\cos(mt')$.
We note that such a quantity, just as  
(\ref{Q1}), does not explicitly 
depend on the 
details of  $f(\OM{k})$ apart from  
$f(0)=1$.

While the asymptotic ($r'\to\infty$) form  of the 
electric field is the same  for different choices of
$f(\OM{k})$, the electric field not too far away 
from the center  of the
studied field configuration is necessarily 
explicitly $f(\OM{k})$-dependent.
It is so because
$f(\OM{k})$  removes the Coulomb singularity
to ensure finiteness of the energy,
which can be achieved in many different ways.
Thereby, to proceed with the discussion 
of physics associated with (\ref{Vprim}),
one must choose  particular $f(\OM{k})$ or 
a class of such functions parametrized 
in some way. We settle for the latter option 
choosing\footnote{Positive even $\beta$'s are preferable because 
      such a choice  facilitates analytical evaluation of 
      the discussed fields.
      The chosen values of $\beta$ guarantee 
      differentiability of the  vector field in the sense 
      specified below (\ref{Vprim}).
      These technical remarks follow from the discussion presented 
      in \cite{BDPeriodic1}.
      }
\be
f(\OM{k})=\B{\frac{m}{\VAREPS{k}}}^\beta, \
\beta=6,8,10,\cdots.
\label{fbeta}
\ee
In such a case 
the energy is 
\be
m\frac{q^2}{8\pi^{3/2}}\frac{\Gamma(\beta-3/2)}{\Gamma(\beta-1)}
\ee
and we can use  the results of 
\cite{BDPeriodic1} to verify that for $r'>|t'|$ 
\begin{align}
\label{Vprimrt}
&V'(t',\BD{r}')\simeq \B{0,-\frac{q\hat{\BD{r}}'}{4\pi r'^2}\frac{\sin(mt')}{m}},\\
\label{CoulT}
&\BD{E}'(t',\BD{r}')\simeq \frac{q\hat{\BD{r}}'}{4\pi r'^2}\cos(mt'),
\end{align}
where   the symbol 
$\simeq$ indicates 
that we have omitted the $\beta$-dependent 
terms because they are $\propto\exp(-m r')$
(such  terms are discussed  in 
Appendix \ref{Subleading_app}). 
Two remarks are in order now.

First,   (\ref{CoulT}) shows that  the
periodically oscillating 
Coulomb component dominates the electric 
field not only for $r'\to\infty$ but 
for $r'>|t'|$ as long as $r'\gg1/m$.
Such a property  should 
not be taken for granted 
because for $r'<|t'|$ the dynamics of 
the discussed   field configuration 
is much more complicated 
and non-universal in the sense that it is 
$\beta$-dependent, which can be inferred from \cite{BDPeriodic1}.
We focus on the universal features in the main body
of this work.

Second, we would like to clarify one point
that was not touched upon 
in \cite{BDPeriodic1}.
Namely, given the fact that (\ref{CoulT}) describes 
 the periodically oscillating  Coulomb field, 
 one may wonder why the vector field 
\be
\tilde{V}'(t',\BD{r}')\simeq \B{\frac{q}{4\pi r'}\cos(m t'),\0}
\label{VprimCC}
 \ee
is not 
describing our field configuration for  
$r'>|t'|$ and $r'\gg1/m$.
Indeed, such a vector field also yields
 (\ref{Bprim}) and 
(\ref{CoulT}) as well as  it presents 
the straightforward generalization 
of the standard textbook  expression 
to the periodically oscillating 
system.
The problem with (\ref{VprimCC}) is that 
it breaks one of  Proca field equations
 in the $r'>|t'|$ region,
where it leads to  $\Box'\tilde{V}' = -m^2\tilde{V}'$
but 
\be
\partial'\cdot\tilde{V}'=
-\frac{q}{4\pi r'}m\sin(mt')\neq0.
\ee
This brings us to the point that the vector 
field is  gauge fixed in the 
Proca theory \cite{Nieto_RMP1971}, 
which one should keep in mind 
while reading this work.
Such an observation is justified 
as follows. Suppose that both $V'$ and 
$\tilde{V}'=V'+(\partial'^\mu\chi')$ 
satisfy Proca field equations, where $\chi'$ is a real-valued function.
Then from $\partial'\cdot\tilde{V}'=0$  one obtains
$\Box'\chi'=0$. By combining this result with 
$\Box'\tilde{V}'=-m^2\tilde{V}'$,
one finds that  $\partial'^\mu\chi'$ must vanish  as long as $m\neq0$.
We mention in passing that there also exists 
the gauge invariant version of the 
Proca theory (see
e.g. \cite{Guralnik1968,Pimentel2015,BDGaugeinv1}).

\section{Another reference frame: general considerations}
\label{Another1_sec}
We will discuss now how the periodically oscillating 
Coulomb field is seen in 
the unprimed reference frame, which 
is moving with velocity $-\BD{v}$
with respect to the primed 
 frame
considered in Sec. \ref{Introduction_sec}. 
The spacetime coordinates in the two systems 
are linked via the 
Lorentz transformation \cite{DraganBook}
\begin{subequations}
\begin{align}
&t'=\gamma(t- \BD{r}\cdot \BD{v}),\\
&\BD{r}'=\BD{r} -\gamma t\BD{v}+
(\gamma-1)\B{\BD{r}\cdot\hat{\BD{v}}}\hat{\BD{v}},
\end{align}
\label{Lorentz}%
\end{subequations}
where $\BD{r}=(x,y,z)$,   $\gamma=1/\sqrt{1-v^2}$, and $v=|\BD{v}|$.

Moreover, the inverse transformation yields the vector field in the 
unprimed reference frame  \cite{MaggioreBook,DraganBook}
\begin{subequations}
\begin{align}
\label{V000}
&V^0(t,\BD{r})=\gamma \BB{V'^0(t',\BD{r}')+\BD{V}'(t',\BD{r}')\cdot\BD{v}},\\
&\BD{V}(t,\BD{r})=
\BD{V}'(t',\BD{r}')    +  \gamma V'^0(t',\BD{r}')\BD{v}
+ (\gamma-1)\BB{\BD{V}'(t',\BD{r}')\cdot\hat{\BD{v}}}\hat{\BD{v}},
\end{align}
\label{V}%
\end{subequations}
where the right-hand sides of these equations are supposed to be evaluated 
via (\ref{Vprim}) combined with (\ref{Lorentz}).
Such obtained vector field $V$ satisfies the unprimed version of 
(\ref{ProcaF}). Thereby, it presents the valid solution of the 
Proca field equations.

The electromagnetic field can be now computed 
from (\ref{V}).  The resulting 
expressions, however, are not insightful 
in their general $f(\OM{k})$-dependent form.
Thereby, we do not list them below.
Still, some results can be obtained without 
assuming a specific form of $f(\OM{k})$. 

First,  the 
electric and magnetic fields 
satisfy 
$\BD{B}(t,\BD{r})=\BD{v}\times\BD{E}(t,\BD{r})$
and
$\BBB{\BD{E}(t,\BD{r})}^2/\gamma^2+\BB{\BD{v}\cdot\BD{E}(t,\BD{r})}^2=
\BBB{\BD{E}'(t',\BD{r}')}^2$.
These expressions can be obtained by 
combining (\ref{Bprim}) with  the following transformation 
law  \cite{MaggioreBook,DraganBook}
\begin{align}
\label{E3d}
&\BD{E}(t,\BD{r})=
\gamma\BB{\BD{E}'(t',\BD{r}')-\BD{v}\times\BD{B}'(t',\BD{r}')}
- (\gamma-1)\BB{\BD{E}'(t',\BD{r}')\cdot\hat{\BD{v}}}\hat{\BD{v}},\\
&\BD{B}(t,\BD{r})=
\gamma\BB{\BD{B}'(t',\BD{r}')+\BD{v}\times\BD{E}'(t',\BD{r}')}
- (\gamma-1)\BB{\BD{B}'(t',\BD{r}')\cdot\hat{\BD{v}}}\hat{\BD{v}}.
\label{B3d}
\end{align}

Second, the charge $Q(t, \RR^3)$, given by the
unprimed version of (\ref{Q1}), can be 
discussed.  As its value  
 can be 
straightforwardly obtained via
the integration of 
$-m^2 V^0=\bnabla\cdot\BD{E}$ (\ref{MoMo}), we
present it as a function of 
\begin{align}
\label{QQ1}
&Q_1(t)  =- m^2\gamma\int d^3r V'^0(t',\BD{r}') 
=q\gamma^2f(\gamma m v)\cos(mt),\\
&
Q_2(t)=- m^2\gamma\int d^3r \BD{V}'(t',\BD{r}')\cdot\BD{v}=-q\gamma^2f(\gamma m v)\cos(mt),
\label{QQ2}
\end{align}
where $\BD{v}\neq\0$ has been assumed and the above expressions 
have been computed  via the following formula 
\be
\int \frac{d^3r}{(2\pi)^3}  
\exp(\ii\BD{k}\cdot\BD{r}'\pm\ii\VAREPS{k}t')=
\gamma \delta(\BD{k}\mp \gamma m \BD{v})\exp(\pm\ii m t)
\label{deltaprim}
\ee
after reversing the order of spatial ($d^3r$) and 
momentum ($d^3k$) integrations; $t'$ and $\BD{r}'$
have been replaced by (\ref{Lorentz}) during the 
evaluation of (\ref{QQ1})--(\ref{deltaprim}).
In terms of $Q_{1,2}(t)$, we find that 
\begin{align}
\label{Qvinfty}%
&Q(t, \RR^3)= Q_1(t) + Q_2(t)=0  \for  \BD{v} \neq \0, \\
&Q(t, \RR^3)= \left.Q_1(t)\right|_{\BD{v}=\0}=q\cos(mt)  \for  \BD{v}= \0,
\label{Qvinfty1}%
\end{align}
where generality of (\ref{Qvinfty})  is seen via the 
observation that 
such a result  does not explicitly depend on $f(\OM{k})$.
(\ref{Qvinfty}) and (\ref{Qvinfty1})  are   rather surprising
because they  show that  $Q(t, \RR^3)$
is discontinuous at $\BD{v}= \0$,
which is not  evident from 
(\ref{V}) or  (\ref{E3d}).
The technical reason for the discontinuity 
is  the following. For $\BD{v}=\0$, 
there is only one contribution to 
$Q(t, \RR^3)$, which coincides with
$\lim_{\BD{v}\to\0}Q_1(t)$.
For $\BD{v}\neq\0$, there are two contributions
to $Q(t, \RR^3)$: 
$Q_1(t)$ and $Q_2(t)$. The latter one does 
not disappear for $\BD{v}\to\0$
despite the appearances. Namely, 
the $\BD{v}$ ``factor''  under the integral in
(\ref{QQ2}), which suggests vanishing of $Q_2(t)$
in such a limit, 
gets canceled during the calculations 
because  it appears in the 
following setting
\be
\frac{\BD{k}\cdot\BD{v}}{\OM{k}^2}\delta(\BD{k}\mp \gamma m \BD{v})
=\pm\frac{\delta(\BD{k}\mp \gamma m \BD{v})}{\gamma m},
\ee
which can be verified  by combining 
(\ref{Vprimvec}), (\ref{QQ2}), and (\ref{deltaprim}).
Nonvanishing of $Q_2(t)$ for $\BD{v}\to\0$ 
makes $Q(t, \RR^3)$ discontinuous.
The physical reason behind such a feature
will be identified in Sec. \ref{Charge_sub}.

\section{Another reference frame: particular solution}
\label{Particular_sec}
We turn our attention here towards the particular solution
specified by (\ref{fbeta}).
The following results  can be compactly 
presented  after the introduction 
of $\BD{R}=\BD{r}-\BD{v}t$, which leads to 
\be
\BD{r}'=\BD{R}+(\gamma-1)(\BD{R}\cdot\hat{\BD{v}})\hat{\BD{v}},
\ r'=\sqrt{R^2+\gamma^2(\BD{R}\cdot\BD{v})^2}.
\ee
For $r>|t|$, the discussed Proca field configuration is 
characterized by 
\begin{subequations}
\begin{align}
&V^0(t,\BD{r})\simeq V^0(\BD{R})\frac{\sin\B{mt/\gamma-m\gamma\BD{R}\cdot\BD{v}}}{m},\\
&V^0(\BD{R})=-q
\frac{\gamma^2\BD{R}\cdot\BD{v}}{4\pi\BB{R^2+\gamma^2(\BD{R}\cdot\BD{v})^2}^{3/2}},
\end{align}
\label{V0rt}%
\end{subequations}

\begin{subequations}
\begin{align}
&\BD{V}(t,\BD{r})\simeq\BD{V}(\BD{R})\frac{\sin\B{mt/\gamma-m\gamma\BD{R}\cdot\BD{v}}}{m},\\
&\BD{V}(\BD{R})=-q
\frac{\BD{R}+\gamma^2(\BD{R}\cdot\BD{v})\BD{v}}{4\pi\BB{R^2+\gamma^2(\BD{R}\cdot\BD{v})^2}^{3/2}},
\end{align}
\label{Vvecrt}%
\end{subequations}

\begin{subequations}
\begin{align}
\label{Eosc1}
&\BD{E}(t,\BD{r})\simeq \BD{E}(\BD{R})\cos\B{mt/\gamma-m\gamma\BD{R}\cdot\BD{v}},\\
&\BD{E}(\BD{R})=q
\frac{\gamma\BD{R}}{4\pi\BB{R^2+\gamma^2(\BD{R}\cdot\BD{v})^2}^{3/2}},
\label{Eosc2}
\end{align}
\label{Eosc}%
\end{subequations}

\begin{subequations}
\begin{align}
&\BD{B}(t,\BD{r})\simeq\BD{B}(\BD{R}) 
\cos\B{mt/\gamma-m\gamma\BD{R}\cdot\BD{v}},\\
&\BD{B}(\BD{R})=\BD{v}\times\BD{E}(\BD{R}).
\end{align}
\label{Bosc}%
\end{subequations}
Some technical  remarks are in order now.

First, we note that (\ref{V0rt})--(\ref{Bosc})
have been obtained by combining 
(\ref{Bprim}), (\ref{Vprimrt}),
and (\ref{CoulT})
with 
(\ref{V})--(\ref{B3d}).
We also note that the 
subleading
contributions to
(\ref{V0rt})--(\ref{Bosc}) 
are discussed in Appendix
\ref{Subleading_app}.
As is explained in Appendix
\ref{Subleading_app}, they are 
proportional to 
\be
\exp\B{-m\sqrt{R^2+\gamma^2(\BD{R}\cdot\BD{v})^2}}
\ee
and as such negligible in the long-distance limit.
Moreover, we note that the subleading contributions  polynomially 
depend on time. This is a non-universal, i.e. $\beta$-dependent 
feature, which is quite distinct from what one finds 
in (\ref{V0rt})--(\ref{Bosc}).

Second, we observe  that the range of validity of  
 (\ref{V0rt})--(\ref{Bosc})
 follows from the fact that $r'>|t'|$
maps onto $r>|t|$ under the Lorentz transformation.
This is easily seen if one notes that these two inequalities 
are equivalent to $X'\cdot X'<0$ and $X\cdot X<0$, where 
$X'=(t',\BD{r}')$ and $X=(t,\BD{r})$.

Third, we note that  (\ref{V0rt})--(\ref{Bosc})
satisfy the following equations in the $r>|t|$ region
\begin{align}
\label{LoLo}
&\partial_t V^0(t,\BD{r})+\bnabla\cdot \BD{V}(t,\BD{r})=0,\\
\label{MoMo}
&\bnabla\cdot\BD{E}(t,\BD{r})=-m^2V^0(t,\BD{r}), \\
&\bnabla\times\BD{E}(t,\BD{r})=-\partial_t\BD{B}(t,\BD{r}),\\
&\bnabla\cdot\BD{B}(t,\BD{r})=0, \\
&\bnabla\times\BD{B}(t,\BD{r})=-m^2\BD{V}(t,\BD{r})+\partial_t\BD{E}(t,\BD{r}),
\label{BoBo}
\end{align}
where $\bnabla=(\partial_{x},\partial_{y},\partial_{z})$.
These equations  are equivalent to the unprimed version of 
(\ref{ProcaF}). We also note  that the primed version of 
(\ref{LoLo})--(\ref{BoBo}) is satisfied by (\ref{Bprim}),
(\ref{Vprimrt}), and (\ref{CoulT}) in the 
$r'>|t'|$ region.
It is presumably worth to stress 
that all this happens despite the fact that 
(\ref{V0rt})--(\ref{Bosc}) as well
as  (\ref{Vprimrt}) and (\ref{CoulT})
provide approximate expressions, which the 
symbol $\simeq$  indicates.

Fourth,  we observe that the $\BD{R}$-dependence
of  (\ref{V0rt})--(\ref{Bosc}) indicates  that 
the studied  field          
configuration moves as  a whole with velocity 
$\BD{v}$, which is  in accordance with our expectations.

In the following, we take a close look at
the above expressions
(Sec. \ref{Comparision_sub})
and the charge that can be computed out of them
(Sec. \ref{Charge_sub}).

\begin{figure}[t]
\includegraphics[width=\pref\columnwidth,clip=true]{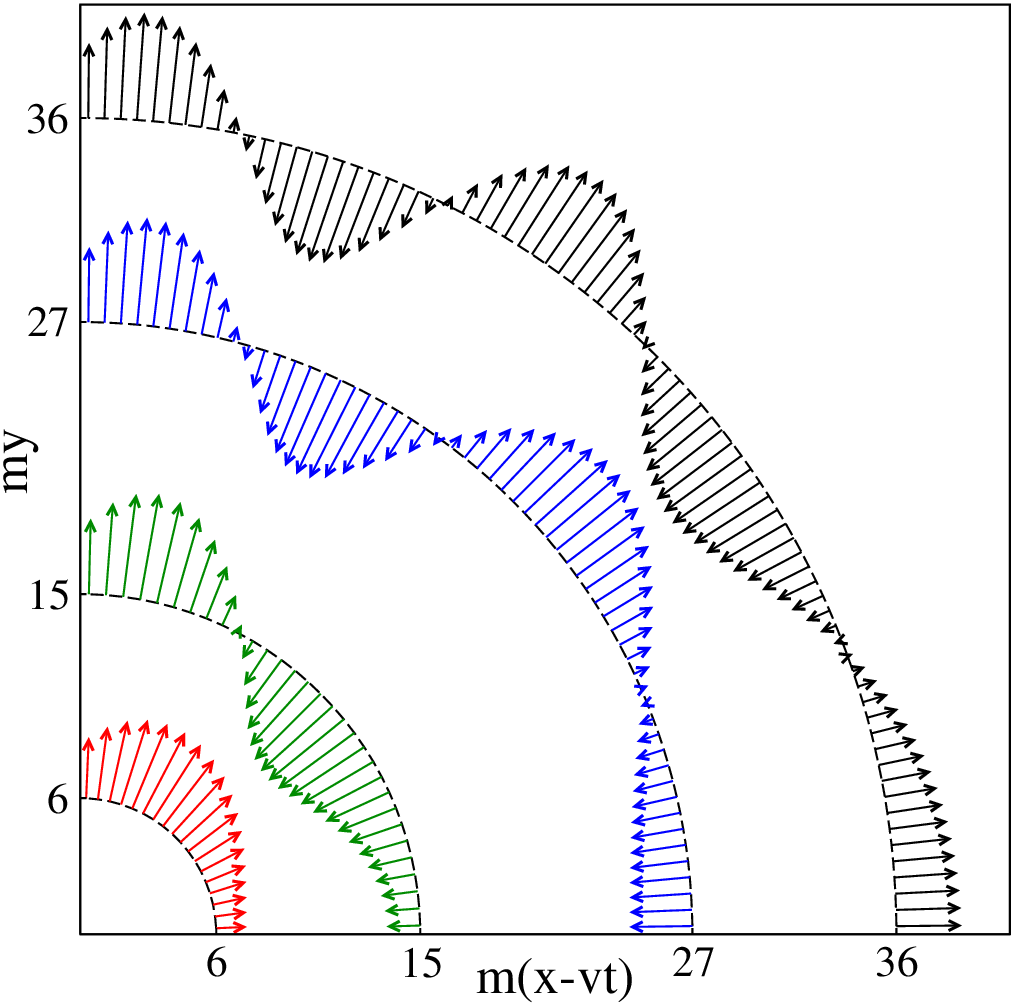} %Eoscil_new3.eps
\caption{The rescaled electric
field in the unprimed reference frame.
Namely, $\BD{E}(t,\BD{r})\times |\BD{r}-\BD{v} t|^2$ in arbitrary units
on the plane $z=0$ for $mt=1$, 
$\BD{v}=(1/3,0,0)$, and 
$mR=m|\BD{r}-\BD{v} t|=6,15, 27,36$. The dashed lines depict 
quarter-circular segments. 
The plotted electric field has 
two components:  leading contribution
 (\ref{Eosc}) and 
subleading contribution (\ref{EEE0})
computed  for $\beta=6$.
The latter contribution rapidly  vanishes as $mR$ grows.
For the  data displayed on the plot,
 mean $|\delta\BD{E}|/|\BD{E}|$ 
is  approximately given by 
$2\cdot10^{-1}$ ($mR=6$),  
$2\cdot10^{-4}$ ($mR=15$),  
$8\cdot10^{-9}$ ($mR=27$),  
and
$2\cdot10^{-12}$ ($mR=36$). The
mean is defined as $\sum_{i=1}^n
      |\delta\BD{E}_i|/|\BD{E}_i|$ divided by $n$,
      where $\BD{E}_i$ is the electric field 
      represented by the 
       $i$-th vector  displayed  
      for a given $mR$ and $\delta\BD{E}_i$ is the subleading 
      contribution to such $\BD{E}_i$ (e.g. $n=18$ for $mR=6$).}
\label{Eoscil_fig}
\end{figure}

\subsection{Comparision to Maxwell theory}
\label{Comparision_sub}

In the $m\to0$ limit,  the electric 
and magnetic fields of the Proca theory,
far away from the center 
of the studied field configuration, 
are described by $\BD{E}(\BD{R})$ and 
$\BD{B}(\BD{R})$:  
the Maxwell theory electromagnetic  field of  the 
charge $q$
moving with the 
constant velocity $\BD{v}$
\cite{Landau}.
Accordingly, vector field (\ref{V0rt}) and 
(\ref{Vvecrt}) for $m\to0$ 
must be related to the 
expressions known in the context 
of the Maxwell theory. This is not immediately 
evident because 
\be
\left(\frac{q}{4\pi }
\frac{\gamma}{\sqrt{R^2+\gamma^2(\BD{R}\cdot\BD{v})^2}},
\frac{q\BD{v}}{4\pi}
\frac{\gamma}{\sqrt{R^2+\gamma^2(\BD{R}\cdot\BD{v})^2}}
\right)
\label{MaxVec}
\ee
is traditionally discussed in the Maxwell theory 
as the vector field  of the  charge moving
with the constant velocity $\BD{v}$ \cite{Landau}.
However, $\lim_{m\to0}(V^\mu)$
and (\ref{MaxVec})  are related via 
the gauge transformation. Indeed,
$\lim_{m\to0}(V^\mu)$
is obtained by subtracting 
$(\partial^\mu\chi)$ from (\ref{MaxVec}),
where 
\be
\chi=\frac{q}{4\pi}\frac{t/\gamma-\gamma\BD{R}\cdot\BD{v}}{\sqrt{R^2+\gamma^2(\BD{R}\cdot\BD{v})^2}}.
\ee

The key differences between (\ref{V0rt})--(\ref{Bosc})
and the  Maxwell theory of 
a  charge moving with a constant velocity 
come from 
the oscillatory terms, which are both time- 
and position-dependent.

The  time dependence seen 
in (\ref{V0rt})--(\ref{Bosc})
implies the 
oscillation period  
\be
\tau=\frac{2\pi}{m\sqrt{1-v^2}}=\gamma\tau',
\ee
which  is dilated with respect to the 
oscillation period in the primed reference frame,
where the center of the studied field configuration 
is at rest. While looking at this result, 
it is instructive 
to realize that 
the periods $\tau$ and $\tau'$
appear in slightly  different
contexts. The period $\tau'$
describes the harmonic oscillations of 
$\BD{V}'$ and  $\BD{E}'$
taking place at any fixed 
point in 
the primed reference frame
that  is far-enough from the center
of the discussed  field configuration.
In such a context, 
the periodically oscillating Coulomb field 
 can be seen 
as  a timekeeping element 
operating in the unit of $\tau'$.
Switching to the unprimed reference frame, 
the period $\tau$ describes the
harmonic oscillations 
of $V$,  $\BD{E}$, and $\BD{B}$
taking place at  the set of spatially-separated 
space-time points: $(t,\BD{R}+\BD{v}t)$, where $\BD{R}$ is a
large-enough constant displacement.

\begin{figure}[t]
\includegraphics[width=\pref\columnwidth,clip=true]{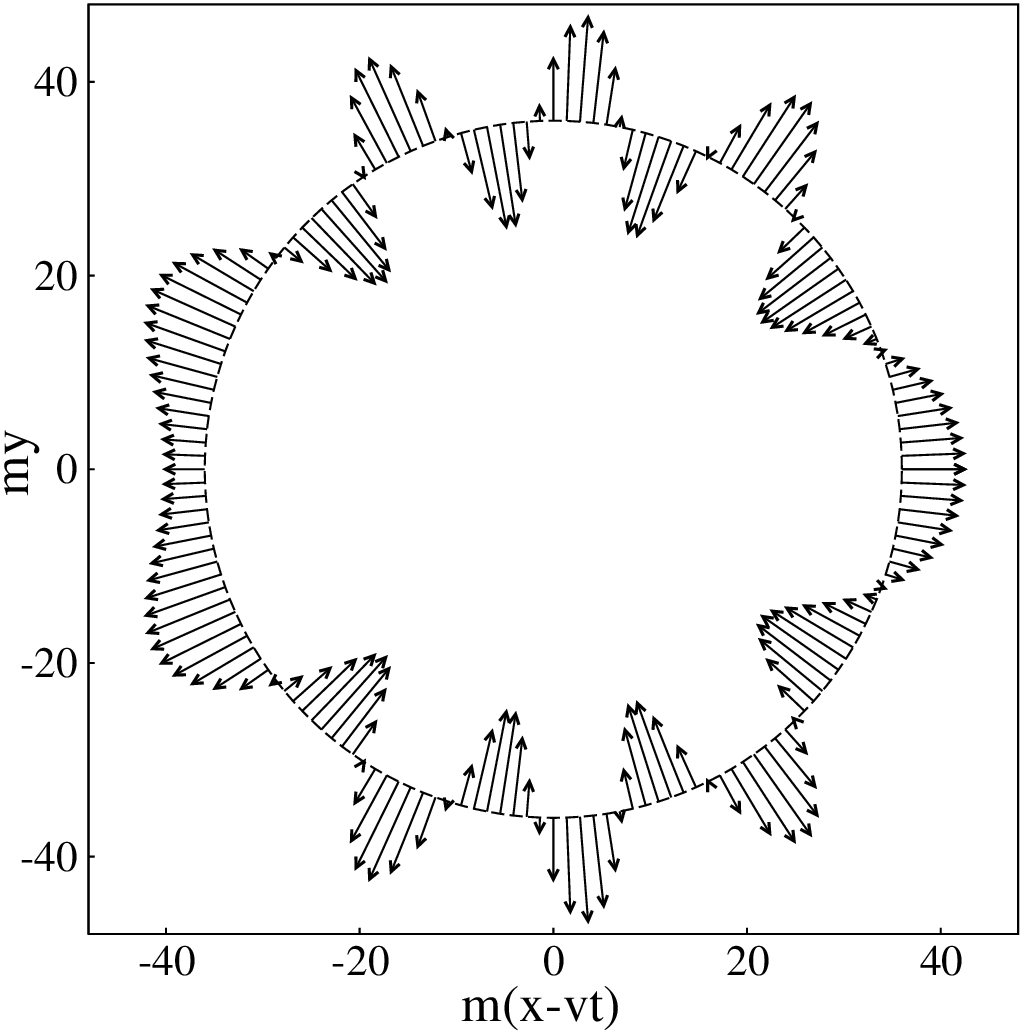} %Eoscil360.eps
\caption{The same as in Fig. \ref{Eoscil_fig} but 
on the whole $x$-$y$ plane and  for $mR=36$ 
only. As  $\BD{E}(t,\BD{r})\propto\BD{R}$, where 
the spatial dependence of the 
proportionality coefficient 
is governed by $R$ and 
$\BD{R}\cdot\BD{v}\propto
\cos[\measuredangle(\BD{R},\BD{v})]$,
one can easily 
visualize  the  three dimensional 
pattern of the  electric field associated 
with this plot.
Such a pattern is obtained by 
rotating the plotted vector field 
about the $y=0$ line.
}
\label{Eoscil360}
\end{figure}

\begin{figure}[t]
\includegraphics[width=\pref\columnwidth,clip=true]{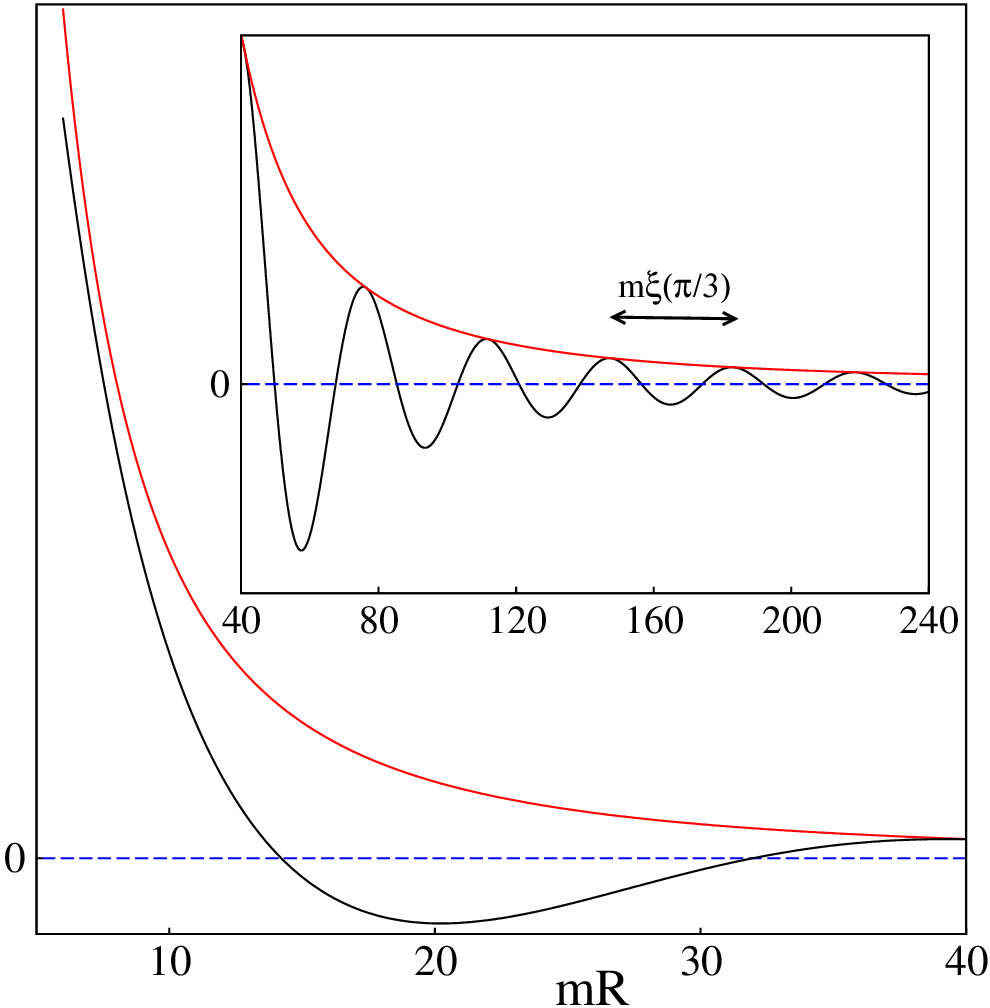} %Etheta1.eps
\caption{The black line depicts 
$\BD{E}(t,\BD{r})\cdot\hat{\BD{R}}$
along the line set by the 
constraint 
$\BD{R}\cdot\BD{v}=Rv\cos(\pi/3)$ 
(leading (\ref{Eosc}) and  subleading 
(\ref{EEE0}) contributions are 
taken into account).
The red line shows the upper envelope of the 
leading contribution to the electric field,
i.e. $\BD{E}(\BD{R})\cdot\hat{\BD{R}}$ 
computed via (\ref{Eosc2}). The blue dashed 
line facilitates  identification of the 
points, where the electric field 
changes direction. The inset
shows the same as the main plot but 
for larger distances.
The plot is prepared for $mt=1$,
$\BD{v}=(1/3,0,0)$, and  $\beta=6$.
There are no quasi-periodic oscillations in 
the main plot, where  $mR$ is at most  of the order of 
$m\xi(\pi/3)\approx36$. The situation is different in the 
inset, where  $mR$ becomes substantially 
larger than $36$. }
\label{Etheta_fig}
\end{figure}

The  position dependence seen in 
the trigonometric functions in 
(\ref{V0rt})--(\ref{Bosc})
adds a
non-trivial local structure to the 
studied field configuration, 
which we illustrate in
Figs. \ref{Eoscil_fig} and \ref{Eoscil360}
in the electric field context.
This can be analyzed  by noting that 
the trigonometric functions
in (\ref{V0rt})--(\ref{Bosc})
are unchanged under the transformation
\begin{subequations}
\be
\BD{R}\to\BD{R}+\xi(\theta)\hat{\BD{R}},
\ \theta=\measuredangle(\BD{R},\BD{v}),
\ee
where 
\be
\xi(\theta) =\frac{2\pi}{mv\cos(\theta)}\sqrt{1-v^2}.
\label{xip}
\ee
\label{xitran}%
\end{subequations}
The quasi-periodic behavior  under (\ref{xitran}) 
is exhibited by  (\ref{V0rt})--(\ref{Bosc}) when $R\gg\xi(\theta)$
because  such a condition guarantees that   $V(\BD{R})$, 
$\BD{E}(\BD{R})$, and $\BD{B}(\BD{R})$ negligibly 
change when (\ref{xitran}) is performed
(see Fig. \ref{Etheta_fig} and comments therein).
All in all, we note that
the  intriguing  pattern is encoded in
(\ref{V0rt})--(\ref{Bosc}).

\subsection{Charge}
\label{Charge_sub}

We begin by defining 
\be
Q(t, {\cal V})=\int_{\cal V} d^3r \bnabla\cdot\BD{E}(t,\BD{r}),
\label{Qvol}
\ee
which conforms to 
previously introduced expression
(\ref{Q1}).
To get quantitative insights into  (\ref{Qvinfty}), we choose 
the integration volume in (\ref{Qvol})
to  be the ball of radius $R$ centered at $\BD{v}t$, where 
$R>(1+v)t$. 
We denote such a ball as ${\cal B}(R)$. 
Its surface  lies inside the region,
where (\ref{Eosc}) holds. Thereby,  we proceed
by replacing the volume integral by the surface one,
which leads to  
\be
Q(t,{\cal B}(R)) \simeq q\gamma\cos(mt/\gamma)
\int_0^1 \mathit{ds} \frac{\cos(p R s)}{\BB{1+(p s/m)^2}^{3/2}},
\label{QBr}
\ee
where $p=\gamma m v$
and $\simeq$ reminds us that expression (\ref{Eosc}), 
which becomes  exact  in the limit of 
$mR\to\infty$,  has been used in order 
to derive  (\ref{QBr}).
For $\BD{v}=\0$, (\ref{QBr}) instantly yields 
the expected   result.
For $\BD{v}\neq\0$,
\be
Q(t, {\cal B}(R))
\simeq \frac{q}{\gamma^2}\cos(mt/\gamma)\frac{\sin(pR)}{pR} +
O\B{\frac{1}{(mR)^2}},
\label{Qpr}
\ee
which  follows from the computations  presented in 
Appendix \ref{Integral_app}.
Taking the limit of $mR\to\infty$,  (\ref{Qvinfty}) 
is obtained.

Alternatively, one may arrive at the same conclusion by 
evaluating  (\ref{Qvol}) over a volume inside a
spheroid given by the constraint
\be
R^2+\gamma^2(\BD{R}\cdot\BD{v})^2= \text{const}.
\label{spher}
\ee
The geometry  of the surface given by the above
equation reflects the Lorentz
contraction in the direction of $\BD{v}$. 
To verify such an observation, we introduce 
$\BD{R}_\perp \perp \BD{v}$
and
$\BD{R}_\parallel \parallel \BD{v}$
such that
$\BD{R}_\perp + \BD{R}_\parallel=\BD{R}$
and write (\ref{spher}) as 
\be
\frac{\BD{R}_\perp\cdot\BD{R}_\perp}{(\gamma\tilde{R})^2} +
\frac{\BD{R}_\parallel\cdot\BD{R}_\parallel}{\tilde{R}^2}=1,
\label{sphEE}
\ee
where we have assumed that the constant in (\ref{spher}) is equal to $(\gamma\tilde{R})^2$.
Such a choice of the right-hand
side of  (\ref{spher})
implies that $\tilde R$ will be  lower-bounded 
in the following calculations  
in  the same way as $R$ in the studies  of 
 $  Q(t, {\cal B}(R)) $. Namely, 
$\tilde{R}>(1+v)t$ ensures  that the spheroid 
lies inside the region of validity of (\ref{Eosc}).
Then, we set $\cal V={\cal S}(\tilde R)$, 
where ${\cal S}(\tilde R)$ is the region of space 
enclosed by (\ref{sphEE}),
and 
replace the volume integral in
(\ref{Qvol}) by the integral  over the
surface given by (\ref{sphEE}). Proceeding 
in the standard way, we  obtain the following result  
without encountering any problems with the evaluation of
the  integral
\be
Q(t,{\cal S}(\tilde R))\simeq 
q\cos(mt/\gamma)\frac{\sin(p\tilde{R})}{p\tilde{R}},
\label{QQpr}
\ee
where $\simeq$ is used in the same context as in (\ref{QBr}).
The limit of $m\tilde{R}\to\infty$ reproduces  (\ref{Qvinfty}).

The disappearance of the 
charge $Q(t,\RR^3)$ for  $\BD{v}\neq\0$ 
can be seen as 
the  result of  self-averaging to zero  of the
 Coulomb-like component of the electric field
over large distances. This  is obviously  caused by 
the spatial dependence of the 
cosine function in (\ref{Eosc}).
Thereby,  
spatial oscillations   of the
electric field provide  the physical reason 
for the discontinuity of $Q(t,\RR^3)$ at $\BD{v}=\0$.

\section{Summary}

There exist  field configurations in the Proca theory,
where the electric field has the  periodically oscillating
Coulomb component. 
This  unusual possibility,
having  no counterpart in the traditional Maxwell theory,
was recently discussed in 
\cite{BDPeriodic1}. In the current work, we 
have described how such  field configurations are seen in
different reference frames.
This has lead to the following  new observations.

First, the periodically oscillating Coulomb field, 
when observed in  a reference frame 
in which it moves as a whole,
exhibits  quasi-periodic 
spatial oscillations. Quite interestingly,
the resulting 
pattern of the electric field in the Proca theory, which is  
depicted in Figs. \ref{Eoscil_fig}--\ref{Etheta_fig},  
does not resemble the well-known 
pattern of the electric 
field of the moving charge
in the Maxwell theory
(similar remarks apply to the 
magnetic field and the vector 
field obtained in our studies of the Proca theory).

Second, the charge associated with  the 
periodically oscillating Coulomb
field--defined 
via  the surface integral 
of the electric field at spatial 
infinity--is  
surprisingly  sensitive 
to the choice of the reference frame.
Namely,   (i) it  periodically oscillates
in  the reference frame in which the 
``center'' 
of the studied field configuration is at rest
and (ii) it  vanishes in all other reference frames.
These observations trigger two remarks.
First, as only the (i) result  
was described in \cite{BDPeriodic1}, 
the present work fills an important 
gap.
Second, the (ii) result  follows from 
the above-mentioned spatial oscillations 
of the electric field.

Finally, we mention that  the
dipole-charged field 
configurations, 
where the  periodic oscillations of either magnetic or 
electric dipole moment are taking place, 
were  recently studied in the quantum 
version of the Proca theory \cite{BDDipole1}.
The reference frame dependence of their 
classical counterparts can be 
 analyzed  akin to what we have presented in this 
work. In fact, one may show   that 
the same trigonometric
factors as in (\ref{V0rt})--(\ref{Bosc})
are responsible for  the spatial 
oscillations of  such 
Lorentz-transformed 
dipole-charged fields  in Proca theory.

\section*{ACKNOWLEDGMENTS}
I am indebted to Bogdan S. Damski
for stimulating comments about this work. 
These studies have
been  supported by the Polish National
Science Centre (NCN) Grant No. 2019/35/B/ST2/00034.
The research for this publication has been also supported 
by a grant from the Priority Research Area DigiWorld under
the Strategic Programme Excellence Initiative at Jagiellonian University.

\appendix
\section{Conventions}
\label{Conv_app}

We adopt  the Heaviside-Lorentz system of units and set $\hbar=c=1$.
Greek and Latin indices of tensors  take values $0,1,2,3$ and   $1,2,3$,
respectively. 
The metric signature is $(+---)$.
$3$-vectors are written in bold, e.g. $V=(V^\mu)=(V^0,\V)$.
We use the Einstein summation convention,  
$(X_{\mu\cdots})^2=X_{\mu\cdots}X^{\mu\cdots}$.
$\Gamma(x)$ is the  gamma function, 
$\hat{\BD{x}}=\BD{x}/|\BD{x}|$,  
$\measuredangle(\BD{x},\BD{y})$ is the angle between 
$3$-vectors $\BD{x}$ and $\BD{y}$,
and $\partial_X=\partial/\partial X$.

\section{Subleading contributions}
\label{Subleading_app}

We provide here  subleading contributions 
to the vector field and the electromagnetic  field 
in both reference frames discussed in the main 
body of this work. We denote them with the symbol
$\delta$ (e.g. 
$\delta\BD{E}'$ ($\delta\BD{E}$) is the subleading contribution
to the electric field in the primed (unprimed) reference frame).

To write the discussed  expressions in the compact form, 
we introduce
\begin{align}
&F_\beta(a,b)=(1+a-a\partial_a)P_\beta(a,b),\\
&\tilde{F}_\beta(a,b)= (1+a-a\partial_a)\partial_b P_\beta(a,b),\\
&G_\beta(a,b)=(1-2\partial_a+\partial_a^2)P_\beta(a,b),
\end{align}
where $P_\beta(a,b)$ are the  bivariate polynomials
whose properties are discussed in
\cite{BDPeriodic1,BDDipole1}.
In particular, 
\be
P_6(a,b)=1+\frac{5a}{8}
+\frac{a^2}{8}
-\frac{b^2}{2}
-\frac{ab^2}{4}
+\frac{b^4}{24},
\label{P66}
\ee
\be
P_8(a,b)=1 + \frac{11 a}{16} + \frac{3 a^2}{16} +
\frac{a^3}{48} - \frac{b^2}{2} - \frac{5 a b^2}{16} 
- \frac{a^2 b^2}{16} + \frac{b^4}{24} + \frac{a b^4}{48} -
\frac{b^6}{720},
\label{P88}
\ee

\begin{align}
P_{10}(a,b)  &=
1 + \frac{93 a}{128} + \frac{29 a^2}{128} + \frac{7 a^3}{192} + \frac{a^4}{384}
- \frac{b^2}{2} - 
\frac{11 a b^2}{32} \nonumber \\& - \frac{3 a^2 b^2}{32} - \frac{a^3 b^2}{96} +
\frac{b^4}{24} + 
\frac{ 5 a b^4}{192} + \frac{a^2 b^4}{192} - \frac{b^6}{720} - 
\frac{a b^6}{1440} + \frac{b^8}{40320}.
\label{P1010}
\end{align}

The subleading contributions to 
 (\ref{Vprimrt}) and  (\ref{CoulT}) read
\begin{align}
\label{ddV01}
&\delta V'^0(t',\BD{r}')=
-\frac{q \exp(-mr')}{4\pi r'}G_\beta(mr',mt'),\\
&
\delta\BD{V}'(t',\BD{r}')=
-\frac{q\rhat' \exp(-mr')}{4\pi r'^2}\frac{\tilde{F}_\beta(mr',mt')}{m},\\
&\delta \BD{E}'(t',\BD{r}')=  
-\frac{q\rhat' \exp(-mr')}{4\pi r'^2}F_\beta(mr',mt').
\label{ddEE1}
\end{align}
(\ref{ddV01})--(\ref{ddEE1}) should be understood as follows.
If we add  $(\delta V'^0,\delta\BD{V}')$ 
to the right-hand 
side of (\ref{Vprimrt}), we will get,
in the $r'>|t'|$ region, an exact result for 
$V'$. Similarly, the addition of 
$\delta\BD{E}'$ to the right-hand 
side of (\ref{CoulT}) yields an exact result
for $\BD{E}'$ in the above-mentioned region.
Finally, we note that $\delta\BD{B}'(t',\BD{r}')=\0$ 
because the magnetic field identically vanishes in 
the primed reference frame.

The subleading contributions to (\ref{V0rt})--(\ref{Bosc}) read
\begin{align}
\delta V^0(t,\BD{r})=
&-q\frac{\gamma^2\exp\B{-m\sqrt{R^2+\gamma^2(\BD{R}\cdot\BD{v})^2}} \BD{R}\cdot\BD{v}}{4\pi m\BB{R^2+\gamma^2(\BD{R}\cdot\BD{v})^2}^{3/2}}
  \tilde{F}_\beta\B{m\sqrt{R^2+\gamma^2(\BD{R}\cdot\BD{v})^2},mt/\gamma-m\gamma\BD{R}\cdot\BD{v}} \nonumber \\
&-q\frac{\gamma\exp\B{-m\sqrt{R^2+\gamma^2(\BD{R}\cdot\BD{v})^2}}}{4\pi\sqrt{R^2+\gamma^2(\BD{R}\cdot\BD{v})^2}}
  G_\beta\B{m\sqrt{R^2+\gamma^2(\BD{R}\cdot\BD{v})^2},mt/\gamma-m\gamma\BD{R}\cdot\BD{v}},
\label{VVV0}
\end{align}
\begin{align}
\delta\BD{V}(t,\BD{r})=&
-q\frac{
\exp\B{-m\sqrt{R^2+\gamma^2(\BD{R}\cdot\BD{v})^2}}
[\BD{R}+\gamma^2(\BD{R}\cdot\BD{v})\BD{v}]
}{4\pi m\BB{R^2+\gamma^2(\BD{R}\cdot\BD{v})^2}^{3/2}}
\tilde{F}_\beta\B{m\sqrt{R^2+\gamma^2(\BD{R}\cdot\BD{v})^2},mt/\gamma-m\gamma\BD{R}\cdot\BD{v}}  \nonumber \\
&-q\frac{\gamma\exp\B{-m\sqrt{R^2+\gamma^2(\BD{R}\cdot\BD{v})^2}}\BD{v} }{4\pi\sqrt{R^2+\gamma^2(\BD{R}\cdot\BD{v})^2}}
 G_\beta\B{m\sqrt{R^2+\gamma^2(\BD{R}\cdot\BD{v})^2},mt/\gamma-m\gamma\BD{R}\cdot\BD{v}},
\label{VVVvec}
\end{align}
\be
\delta\BD{E}(t,\BD{r})=
-q\frac{\gamma\exp\B{-m\sqrt{R^2+\gamma^2(\BD{R}\cdot\BD{v})^2}}\BD{R} }{4\pi\BB{R^2+\gamma^2(\BD{R}\cdot\BD{v})^2}^{3/2}}
 F_\beta\B{m\sqrt{R^2+\gamma^2(\BD{R}\cdot\BD{v})^2},mt/\gamma-m\gamma\BD{R}\cdot\BD{v}},
\label{EEE0}
\ee
\be
\delta\BD{B}(t,\BD{r})=\BD{v}\times\delta\BD{E}(t,\BD{r}).
\label{BBB0}
\ee
Several remarks are in order now.

First, results (\ref{ddV01})--(\ref{ddEE1})
can be found in
\cite{BDPeriodic1}. 
They are gathered here
for the sake of completeness of our discussion.
Results (\ref{VVV0})--(\ref{BBB0})
have been obtained 
from (\ref{ddV01})--(\ref{ddEE1}) via 
(\ref{V})--(\ref{B3d}).
If we add (\ref{VVV0}), (\ref{VVVvec}), (\ref{EEE0}), and
(\ref{BBB0}) to (\ref{V0rt}), (\ref{Vvecrt}), (\ref{Eosc}), and 
(\ref{Bosc}),
respectively, we will get  exact expressions for $V^0$, $\BD{V}$, 
$\BD{E}$, and $\BD{B}$   in the $r>|t|$ region.

Second, the above expressions show that  
 subleading contributions
exponentially vanish with the distance from the 
center of the studied field configuration.
This happens  in both 
reference frames.
The distance at which the subleading contributions 
become  negligible depends on the parameter $\beta$.

Third, we note that  (\ref{Eosc}) and 
$\delta\BD{E}(t,\BD{r})$ are proportional to $\BD{R}$, where 
the spatial dependence of the proportionality coefficient 
is determined by $R$ and 
$\BD{R}\cdot\BD{v}$. This should be kept in mind when one 
looks at Figs. \ref{Eoscil_fig}--\ref{Etheta_fig}.

Fourth,  we note that
in the $r>|t|$ region
(\ref{VVV0})--(\ref{BBB0}) 
satisfy (\ref{LoLo})--(\ref{BoBo}), where 
$V^0$, $\BD{V}$,   $\BD{E}$, and $\BD{B}$ are replaced by 
$\delta V^0$, $\delta\BD{V}$,
$\delta\BD{E}$, and $\delta\BD{B}$.
In the same sense, the subleading contributions in the primed reference frame
satisfy the primed version of (\ref{LoLo})--(\ref{BoBo})
 in the $r'>|t'|$ region.

\section{Integral from expression  (\ref{QBr})}
\label{Integral_app}

The integral of interest here is 
\begin{subequations}
\begin{align}
&I=\int_0^1 \mathit{ds}\, g(s) \cos(pR s),\\
&g(s)= \frac{1}{\BB{1+(p s/m)^2}^{3/2}},
\end{align}
\end{subequations}
where $p=\gamma m v$.
Integrating twice by parts, we find 
\begin{subequations}
\begin{align}
&I=\frac{1}{\gamma^3}\frac{\sin(p R)}{p R} + \Delta,\\
&\Delta=\frac{1}{(p R)^2}\int_0^1 \mathit{ds}\, g''(s)\BB{\cos(p R)-\cos(p R s)},
\end{align}
\end{subequations}
where $d^2g/ds^2=g''$.
To bound  $\Delta$, we compute
\begin{align}
|\Delta|&\le \frac{2}{(p R)^2} \int_0^1 \mathit{ds} \BBB{g''(s)}\nonumber \\
&=\frac{6}{(m R)^2} \int_0^1 \mathit{ds} 
\frac{\BBB{1-4(p s/m)^2}}{\BB{1+(p s/m)^2}^{7/2}} \le \frac{6}{(m R)^2} \int_0^1 \mathit{ds} 
\frac{1+4(p s/m)^2}{\BB{1+(p s/m)^2}^{7/2}}  \nonumber \\
&= \B{6+4v^2-\frac{18}{5}v^4}
\frac{1}{\gamma(m R)^2}
\le \frac{64}{9} \frac{1}{\gamma(m R)^2}, 
\end{align}
where  knowledge of the following easy-to-verify
indefinite integral has been  employed 
\be
\int \mathit{ds}
\frac{1+4(p s/m)^2}{\BB{1+(p
s/m)^2}^{7/2}}=\frac{s+\frac{8}{3}(p/m)^2 s^3+\frac{16}{15}(p/m)^4 s^5}{\BB{1+(ps/m)^2}^{5/2}}.
\ee

%\bibliography{../Scommon/reference.bib} 

\end{document}